\documentclass[aps,prl,twocolumn,amsmath,nofootinbib]{revtex4-1}
\usepackage{bm}
\usepackage{mathrsfs}
\usepackage{graphicx}
\usepackage[skip=2pt,font=small]{caption}

\DeclareMathOperator{\tr}{Tr}

\providecommand{\mat}[1]{\ensuremath{\bm{\mathsf{#1}}}}

\begin{document}
\title{Room temperature chiral discrimination in paramagnetic NMR spectroscopy}

\author{Alessandro Soncini}
\email{asoncini@unimelb.edu.au}
\author{Simone Calvello}

\affiliation{School of Chemistry, University of Melbourne, VIC 3010, Australia}

\begin{abstract}
A recently proposed theory of chiral discrimination in NMR spectroscopy based on the detection of a molecular electric polarization $\bm{P}$ rotating in a plane perpendicular to the NMR magnetic field [A.~D.~Buckingham, J. Chem. Phys. {\bf 140}, 011103 (2014)], is here generalized to paramagnetic systems.  Our theory predicts new contributions to $\bm{P}$, varying as the square of the inverse temperature.  Ab initio calculations for ten Dy$^{3+}$ complexes, at 293K, show that in strongly anisotropic paramagnetic molecules $\bm{P}$ can be more than 1000 times larger than in diamagnetic molecules, making \textit{paramagnetic} NMR chiral discrimination amenable to room temperature detection.
\end{abstract}
\maketitle

Despite its central role in biological processes and in a wide range of chemical reactions, chirality, the property of molecules lacking improper symmetry elements to be distinguishable from their mirror image (or enantiomer), remains a challenging property to detect and quantify~\cite{ChiralAnalysis2006}, making the development of new spectroscopic techniques to achieve this goal a strategic research field~\cite{Fischer2000,Buckingham2004,Fischer2005,BuckinghamFischer2006,Li2008,Hiramatsu2012,Patterson2013,Buckingham2014}. 

Magnetic resonance spectroscopies, despite being among the most useful characterization techniques due to their high sensitivity to tiny details of the geometrical and electronic structure of molecules, are blind to chirality.
However, it has been recently proposed by Buckingham~\cite{Buckingham2004,Buckingham2014} and Fischer~\cite{BuckinghamFischer2006} that NMR could be used to achieve chiral discrimination for closed-shell chiral molecules via a minor modification of the experimental set up, so to make it fit for the detection of a rotating average electric polarization $\overline{\bm{P}}$ induced by the combined effect of the NMR magnetic field $\bm{B}$, and the nuclear magnetic dipole moment $\bm{m}^I$ associated to a nucleus $I$ of the chiral molecule, rotating in the plane perpendicular to $\bm{B}$ following a resonant $\pi/2$ radiofrequency pulse.   In particular, the induced $\overline{\bm{P}}$ is always oriented along $\bm{B}\times\bm{m}^I$ (thus rotating with $\bm{m}^I$), points in opposite directions for the two enantiomers (hence its chirality-sensitivity), and is proportional to the pseudoscalar $\sigma^{(1)}$, isotropic average of a third-rank tensor $\sigma_{ijk}$ ($i,j,k=x,y,z$) known as shielding polarizability~\cite{Buckingham2004,BuckinghamFischer2006,Lazzeretti2008}.  However, computational estimates of $\sigma^{(1)}$ in diamagnetic molecules suggest that it is generally too small to be detected~\cite{Monaco2011,Buckingham2014,Buckingham2015}.    

  In this Letter we propose a theory of paramagnetic NMR chiral discrimination which is valid for molecules with a ground state of arbitrary degeneracy.  We describe the response of the degenerate system in terms of a generalized shielding polarizability tensor $\Phi_{ijk}$  defined by analytical third--derivatives of the free energy, reducing to $\sigma_{ijk}$ in the limit of a non-degenerate ground state. 
The proposed free-energy response theory features previously unexplored terms proportional to the square of the inverse temperature $\beta=1/k_BT$, so that $\sigma^{(1)}$ becomes $\sigma^{(1)}+\beta^2 \sigma^{(1p)}$.   Finally, we present ab initio calculations showing that $\beta^2\sigma^{(1p)}$ yields a contribution to $\overline{\bm{P}}$ at 293K that is orders of magnitude larger than in closed-shell molecules, potentially observable at room temperature.

 {\it Theory}.  Let us consider an ensemble of nuclear magnetic moments $\bm{m}^I = \mu_N g_I  \bm{I}$, with $\mu_N$ the nuclear magneton, $g_I $ the g-factor of the probed nucleus $I$, and $\bm{I}$ the nuclear spin in units of $\hbar$.   In a pulsed NMR experiment,  the nuclear moments $\bm{m}^I$,  partially aligned along a magnetic field $\bm{B}=B_0 \bm{u}_z$, are flipped in the $xy$-plane perpendicular to $\bm{B}$ by a radiofrequency $\pi/2$-pulse in resonance with the nuclear Larmor frequency $\omega^I = \mu_N g_I  (1-\sigma) |\bm{B}|$.  Here $\sigma$ is the nuclear shielding constant, the isotropic average of the shielding tensor $\bm{\sigma}$ describing the departure from $\bm{B}$ of the local magnetic field at nucleus $I$ due to surrounding electrons. The coherent in-plane precession of $\bm{m}^I$ gives rise to a free-induction decay (FID) signal picked up by a coil with axis in the $xy$-plane, whose Fourier transform provides the NMR spectrum.  
 
The combined effect of $\bm{B}$ and $\bm{m}^I$ on the electrons of a diamagnetic molecule has been shown to induce an electric dipole polarization~\cite{Buckingham2004,BuckinghamFischer2006,Buckingham2014}, which can be obtained by expanding the energy of the molecule $W( \bm{E},\bm{B},\bm{m}^I)\equiv W$ in a power series of $\bm{B}$, $\bm{m}^I$, and an external electric field $\bm{E}$, thermally averaged over all molecular orientations in the presence of $\bm{B}$ and $\bm{m}^I$~\cite{Buckingham2014}. 
Truncation of the ensuing power series to leading order in the inducing fields yields the average induced polarization $\overline{P_i}\equiv\overline{P}_{d,i}$ as~\cite{Buckingham2004,BuckinghamFischer2006,Buckingham2014}: 
 \begin{equation}
 \label{polariCS}
  \overline{P}_{d,i}  = -\left\langle \frac{\partial W} {\partial E_i} \right\rangle 
\approx -\left\langle \sigma_{ijk} 
+ \beta  \bar{\mu}_i \sigma_{jk}\right\rangle_{\mathrm{rot}} B_j m^I_k  
\end{equation}
where subscript $d$ stands for diamagnetic, $i = x,y,z$ are defined with respect to the laboratory frame, sum over repeated indices is implied, $\bm{\bar{\mu}}$ is the ground state electric dipole moment of the molecule, $\langle\dots\rangle$ stands for thermal average over all orientations, $\langle\dots\rangle_{\mathrm{rot}}$ stands for pure rotational average (i.e. isotropic average), $\sigma_{ij}$ is the nuclear shielding tensor, and $\sigma_{ijk}$  the shielding polarisability tensor defined as an energy third-derivative at zero-fields:
\begin{equation}
\label{enederiv}
\sigma_{ijk} = \left. \frac{\partial^3 W}{\partial E_i \partial B_j \partial m^I_k} \right|_0.
\end{equation}
The pseudoscalar $\sigma^{\left(1\right)}$~\cite{BuckinghamFischer2006,Buckingham2014} 
is evaluated by contraction of $\sigma_{ijk} 
+ \beta  \bar{\mu}_i \sigma_{jk}$ with the Levi-Civita totally antisymmetric tensor $\epsilon_{ijk}$~\cite{thirunamachandran}, leading to $\left\langle \sigma_{ijk} + \beta \bar{\mu}_i \sigma_{jk} \right\rangle_{\mathrm{rot}}  =\sigma^{\left(1\right)}\epsilon_{ijk}$, with 
$\sigma^{\left(1\right)} = \frac{1}{6}\epsilon_{abc}\left(\sigma_{abc} + \beta \bar{\mu}_a \sigma_{bc}\right)$, $a,b$ and $c$ defined with respect to the molecule-fixed frame.

In molecules with a degenerate ground state more than one level is populated after the NMR fields partially split the degeneracy.
 We thus proceed to define a perturbative expansion of the electronic free energy~\cite{FeynmanStatBook} as recently proposed for pNMR chemical shifts~\cite{prl_pnmr,jcpcomm_pnmr,jcpbig_pnmr}, assuming that electrons are in thermal equilibrium with the NMR fields (i.e. electronic spin-lattice relaxation is fast compared to the NMR dynamics). The electronic free energy $F\left(\bm{E},\bm{B},\bm{m}^I \right) \equiv F$  reads:
\begin{equation}
F = -\beta^{-1} \ln\left\langle\tr\rho\right\rangle_{\mathrm{rot}} 
\label{FreeEnergy}
\end{equation}
where $\rho = e^{-\beta H}$ is the equilibrium statistical operator in the presence of the external fields, fully defined by the molecular Hamiltonian:
\begin{equation}
\label{totalham}
H = H_0 + \lambda V_1 + \lambda^2 V_2,
\end{equation}
$\left\langle\tr\rho\right\rangle_{\mathrm{rot}}$ is the rotationally averaged partition function.
Here $H_0$ is the molecular electrostatic Hamiltonian, $V_1$ is the sum of the Zeeman $V_{zee}$, hyperfine $V_{hf}$ and electric dipole/field interaction $V_{el}$ Hamiltonians, and $V_2$ is the term bilinear in $\bm{B}$ and $\bm{m}^I$, responsible for the diamagnetic contribution to the nuclear shielding tensor~\cite{Abragam1961}:
\begin{equation}
\label{ham}
V_1 = V_{zee} + V_{hf} + V_{el} \;\;\; \text{and}  \;\;\;\; V_2=V_{\mathcal{D}}
\end{equation}
with 
\begin{equation}
\label{hamdetails}
\begin{array}{lcl}
V_{zee} & =  &-\bm{\mathcal{M}}\cdot\bm{B} \\
V_{hf}    & = & \bm{\mathcal{F}}\cdot\bm{m}^I \\
V_{el}   & = & -\bm{\mu}\cdot\bm{E}\\
V_{\mathcal{D}}  & = & \bm{B}\cdot\bm{\mathcal{D}}\cdot\bm{m}^I
\end{array}
\end{equation}
where $\bm{\mathcal{M}} = \mu_B\left( \bm{L}+2\bm{S}\right)$ is the electronic magnetic dipole operator ($\mu_B$ the Bohr magneton, in atomic units $\alpha/2$, with $\alpha$ the fine structure constant), $\bm{\mu}$ is the electric dipole operator, $\bm{\mathcal{F}}$ is the hyperfine field induced by the electrons at the site of nucleus $I$, consisting of Fermi contact, spin-dipolar and nuclear spin-electron orbit contributions, and $\bm{\mathcal{D}}$ is the diamagnetic shielding operator.

The average electric polarization $\overline{P}_i$  is thus defined as:
 \begin{equation}
 \label{polariOS}
\overline{P}_i =  \overline{P}_{d,i}+\overline{P}_{p,i} = -\frac{\partial F}{\partial E_i },
 \end{equation}
where $\overline{P}_{p,i}$ is the new paramagnetic contribution to $\overline{P}_i$, while $\overline{P}_{d,i}$ follows from Eq.~(\ref{polariCS}).  
To obtain an analytical expression for Eq.~(\ref{polariOS}) in weak fields, we expand $\rho$ in Eq.~(\ref{FreeEnergy}) in powers of $\lambda$:
\begin{equation}
\label{rhopt}
\rho = \rho_0 + \lambda\rho_1 + \lambda^2\rho_2 + \lambda^3\rho_3+\dots
\end{equation}
where $\rho_0=e^{-\beta H_0}$, and:  
\begin{equation}
\label{rhon}
\rho_n = -\int_{0}^{\beta} d\omega e^{\left(\omega-\beta\right)H_0} \left[  V_1 \rho_{n-1}\left(\omega\right) + V_2 \rho_{n-2}\left(\omega\right)  \right].
\end{equation}
Substituting Eq.~(\ref{rhopt}) into Eq.~(\ref{FreeEnergy}), we obtain:
\begin{equation}
\label{free3}
F  =  F_0+\lambda F_1 + \lambda^2 F_2 + \lambda^3 F_3 +\dots.
\end{equation}

Neglecting contributions to $\bm{P}$ in Eq.~(\ref{polariOS}) that are not bilinear in $\bm{B}$ and $\bm{m}^I$, thus could not describe a rotating polarization in the pulsed NMR experiment, and accounting for rotational averaging in the fast spin-lattice relaxation limit, we obtain to leading order in the fields:
\begin{equation}
 \label{polariOS2}
\overline{P}_i \approx -\frac{\partial F_3}{\partial E_i }= -\Phi_{ijk}B_j m^I_k 
 \end{equation}
where we have introduced the {\it generalized shielding polarizability}  $\Phi_{ijk}$, a third-rank tensor defined by:
\begin{equation}
 \label{shieldingpolari2}
\Phi_{ijk} = \left. \dfrac{\partial^3 F}{\partial E_i \partial B_j \partial m^I_k} \right|_0 =  \dfrac{\partial^3 F_3}{\partial E_i \partial B_j \partial m^I_k},
\end{equation}
and where the relevant free-energy contribution $F_3$ reads:
\begin{equation}
\label{f3}
F_3 = -\beta^{-1} \left\langle \frac{\tr\rho_3}{\tr\rho_0}\right\rangle_{\mathrm{rot}}.
\end{equation}

To evaluate $\Phi_{ijk}$ we thus first need an explicit expression for $F_3$ as function of the NMR fields. To obtain $\tr\rho_3$ we use Eq.~(\ref{rhon}) with $n=3$, leading to:
\begin{equation}
\label{ro3}
\begin{split}
\frac{\tr\rho_3}{\tr\rho_0} & = \frac{\beta}{2} \left\langle \displaystyle\int_0^\beta \mathrm{d}u \; T\left[ V_1(u)V_2(0)+V_2(u)V_1(0)\right] \right\rangle_0  \\
         & -  \frac{\beta}{6} \left\langle \displaystyle\int_0^\beta  \!\!\!\! \displaystyle\int_0^\beta \!\! \mathrm{d}u\mathrm{d}v  \;  
         T\left[V_1(u)V_1(v)V_1(0)\right]  \right\rangle_0,
\end{split}
\end{equation}
where $\left\langle O \right\rangle_0 \equiv \tr(\rho_0 O)/\tr\rho_0$, the imaginary-time Heisenberg picture operators $V_i(\omega)$  (with $i=1,2$) are:
\begin{equation}
\label{heisenbergOP}
V_i(\omega) = e^{\omega H_0} V_i e^{-\omega H_0},
\end{equation}
and $T\left[\dots\right]$ represents the imaginary-time-ordering operator permuting the $V_i(\omega)$'s so that operators with later imaginary-time arguments are shifted to the left.  

We can now evaluate $\Phi_{ijk}$ in Eq.~(\ref{shieldingpolari2}), by substitution of Eq.~(\ref{ro3}) into Eq.~(\ref{f3}), and  by carrying out the third mixed-derivative as prescribed in Eq.~(\ref{shieldingpolari2}). We obtain:

\begin{eqnarray}
\Phi_{ijk} & = & \frac{1}{6} \sum_{P\in S_3}\left\langle \displaystyle\int_0^\beta  \!\!\!\! \displaystyle\int_0^\beta \!\! \mathrm{d}u\mathrm{d}v  \;  
         T\left[P\mu_i(u)\mathcal{M}_j(v)\mathcal{F}_k\right]  \right\rangle_{0,\mathrm{rot}} \nonumber \\
        &  & \;\;\;\;\; + \frac{1}{2} \sum_{P\in S_2}\left\langle \int_0^\beta \!\!\mathrm{d}u \; T\left[ P\mu_i(u)\mathcal{D}_{jk}\right]\right\rangle_{0,\mathrm{rot}}  \label{phi3details}.
\end{eqnarray}
The two terms on the right-hand side of Eq.~(\ref{phi3details}) feature a sum over all permutations of the operators involved, 
with each permutation $P$ switching the operators without moving their imaginary-time arguments.  

Eq.~(\ref{phi3details}) provides a generalization of Buckingham's theory to ground states of arbitrary degeneracy, and hence represents a central result of this Letter.  
Note that we can identify $\Phi_{ijk}$ in Eq.~(\ref{phi3details}) with the shielding polarizability tensor $\sigma_{ijk}$ in Eq.~(\ref{polariCS})~\cite{Buckingham2004,BuckinghamFischer2006,Lazzeretti2008}, however here featuring all contributions, both temperature-independent and temperature-dependent, in a single expression. Temperature dependent contributions were previously discussed~\cite{Buckingham2014} for diamagnetic molecules, as summarized by Eq.~(\ref{polariCS}).  Our result Eq.~(\ref{phi3details}) also includes new contributions that are non-zero only for paramagnetic molecules, essentially arising from the NMR paramagnetic shift $\beta\sigma^p$, whose theory~\cite{Kurland1970,Moon2004,Vaara2008} has been recently generalized to ground states of arbitrary degeneracy~\cite{prl_pnmr,jcpcomm_pnmr,jcpbig_pnmr,AutschbachTemperature2015}.

Explicit sum-over-states expressions for $\Phi_{ijk}$  can be derived using the basis  $H_0 \left| m\mu \right\rangle = \epsilon_m \left| m\mu \right\rangle$, with $\mu = 1, \dots, d_m$ enumerating degenerate states, by introducing the resolution of identity $I=\sum_m \sum_{\mu=1}^{d_m} \left|m\mu\right\rangle\left\langle m\mu\right|$ in between operators in Eq.~(\ref{phi3details}), thus turning the thermal propagators Eq.~(\ref{heisenbergOP}) into exponential functions of the integration variables. 

For the special but common case of a thermally isolated ground state (TIGS), $\left| a\alpha \right\rangle$  ($\alpha = 1, \dots , d_a$), this leads to:
\begin{equation}
\label{PiDeltaAnalysis}
\Phi_{ijk} = \frac{1}{6}\epsilon_{abc}\left(\sigma_{abc} + \beta  \bar{\mu}_a \sigma_{bc} + \beta^2 \sigma^{p}_{abc}\right)\epsilon_{ijk}
\end{equation}
where $\sigma_{ab}$ and $\sigma_{abc}$ in Eq.~(\ref{PiDeltaAnalysis}) are equivalent to the closed-shell magnetic shielding and shielding polarizability tensors in Eq.~(\ref{polariCS}), which are generally small second and third order response properties~\cite{Buckingham2004,BuckinghamFischer2006,Lazzeretti2008}. 
On the other hand, the new paramagnetic contribution in Eqs.~(\ref{PiDeltaAnalysis}) features the third rank tensor $\sigma^p_{abc}$:
\begin{equation}
 \sigma^{p}_{abc}  =  d_a^{-1}\!\!\!\sum_{\alpha\alpha'\alpha''}  \!\!\! \Re \left[ \left\langle a\alpha \right|  \mu_a \left| a\alpha' \right\rangle
 \left\langle a\alpha' \right|  \mathcal{M}_b \left| a\alpha'' \right\rangle   
 \left\langle a\alpha'' \right|\mathcal{F}_c \left| a\alpha \right\rangle \right]  \label{sigmapi} 
 \end{equation} 
Since Eq.~(\ref{sigmapi}) is independent of inverse energy gaps, it is expected to give a sizeable contribution to the chiral polarization  Eq.~(\ref{polariOS2}).  This will be assessed next via ab initio calculations.

 {\it Estimate of the chiral polarization.} It follows immediately from Eqs.~(\ref{polariCS}), (\ref{polariOS2}), (\ref{PiDeltaAnalysis}) and (\ref{sigmapi}) that $\overline{P}_{p,i}$  for an ensemble of tumbling molecules with a TIGS reads: 
 \begin{equation}
 \label{polariPARAM}
 \begin{split}
  \overline{P}_{p,i} = -\beta^2 \sigma^{\left(1p\right)} \left(\bm{B}\times\bm{m}^I\right)_i 
 \end{split}
 \end{equation}
 where $\sigma^{\left(1p\right)} \equiv \frac{1}{6}\epsilon_{abc} \sigma^p_{abc}$.
 Since $\bm{m}^I$ is rotating following a $\pi/2$ pulse, the chiral electric polarization $\overline{P}_{p,i}$ follows such rotation, which can be observed as an AC-voltage induced in a capacitor sandwiching the pick-up coil in a modified NMR/pNMR experiment~\cite{BuckinghamFischer2006,Pelloni2013}.

Our aim in this last part is to provide an estimate of $\beta^2\sigma^{\left(1p\right)}$ at room temperature, and determine the voltage that the associated electric polarization Eq.~(\ref{polariPARAM}) will produce.  We consider here the case of a magnetically anisotropic Kramers Doublet (KD) ground state with a non-zero permanent electric dipole moment. 
As it will become apparent soon, strong magnetic anisotropy is found to be an essential property for a chiral paramagnetic molecule to produce a large rotating electric polarization in a pulsed NMR experiment.
This property is commonly encountered in low-symmetry Ln$^{3+}$ complexes displaying single-molecule magnet (SMM) behaviour~\cite{Winpenny2013}, for which detection of paramagnetic shifts in $^1$H-NMR and $^{13}$C-NMR experiments has proved very useful in the elucidation of their electronic structure~\cite{IshikawaPNMR2003,Enders2015}.  
The ground KD can be associated to a pseudo-spin $\tilde{s}=1/2$, so that the magnetic moment $\mathcal{M}_i$, and hyperfine field $\mathcal{F}_i$ operators, written in terms of the pseudo-spin operator $\tilde{\bm{S}}$, are  $\mathcal{M}_i = -\mu_B g_{ij} \tilde{S}_j$ and $\mathcal{F}_i = \left(g_I\mu_N\right)^{-1} A^I_{ji} \tilde{S}_j$,
where $g_{ij}$  and $A^I_{ji}$ are the Zeeman $g$-tensor and hyperfine $A$--tensor, respectively, collecting the matrix elements of $\bm{\mathcal{M}}$ and $\bm{\mathcal{F}}$  in the basis of e.g. ab initio KD wavefunctions.  The representation of the electric moment component $\mu_i$ in the KD basis is forced by time-reversal symmetry to be a scalar $\bar{\mu}_i$ times the $2\times2$ identity matrix.
Within this pseudo-spin representation~\cite{prl_pnmr}, 
we obtain:
 \begin{equation}
 \label{polariPARAMexplicit}
 \begin{split}
  \sigma^{\left(1p\right)} = & - \frac{\mu_B}{24 g_I \mu_N} \epsilon_{abc} \bar{\mu}_a  g_{bd}A^I_{dc}.
 \end{split}
 \end{equation}
Since $\sigma^{\left(1p\right)}$ Eq.~(\ref{polariPARAMexplicit}) is now expressed in terms of the well-known response tensors  $g_{ik}$ and $A^I_{kj}$, we can proceed with its evaluation via ab initio methods. We choose here to avoid accurate evaluation of $A^I_{ij}$, which is a non trivial task for a KD with strong multiconfigurational character~\cite{AutschbachHyperfine2012,AutshbachRAShyperfine2015}, and invoke instead the well-known dipolar approximation~\cite{Kurland1970} expressing $A^I_{ij}$ as the dipolar magnetic field induced at the site of the probed nucleus $I$ by the paramagnetic ion, making it proportional to the g--tensor itself and the inverse cube of distance $\bm{R}_I$ between the paramagnetic ion and the probed nucleus $I$:
 \begin{equation}
 \label{Adipolar}
A^I_{ij}\approx  \frac{\mu_B \mu_N g_I}{R_I^5} g_{ik} \left( \delta_{kj} R_I^2  -  3 R_{I,k}R_{I,j}\right).
 \end{equation}
Introducing polar coordinates for nucleus $I$ with respect to the reference frame in which the g-tensor is diagonal $\mat{g} = \mathrm{diag}\left(g_{xx},g_{yy},g_{zz}\right)$ (principal axes frame), $X_I = R_I \sin\chi \cos\Omega$, $Y_I = R_I \sin\chi \sin\Omega$, and $Z_I = R_I \cos\chi$, and substituting Eq.~(\ref{Adipolar}) into Eq.~(\ref{polariPARAMexplicit}), we obtain:
\begin{eqnarray}
\label{sigma1dipolar}
\sigma^{\left(1p\right)} & = & \frac{\mu_B^2}{8}\frac{\sin\chi}{R_I^3}\left[ 
\overline{\mu}_x\left(g_{yy}^2-g_{zz}^2\right)\sin\Omega\cos\chi \right.  \nonumber \\
& & +  \overline{\mu}_y\left(g_{zz}^2-g_{xx}^2\right)\cos\Omega\cos\chi \nonumber \\
& & +  \left. \overline{\mu}_z\left(g_{xx}^2-g_{yy}^2\right)\sin\chi\frac{\sin2\Omega}{2}  
\right]
\end{eqnarray}
Eq.~(\ref{sigma1dipolar}) exposes the fundamental role that magnetic anisotropy plays in the physics predicted here.  The pseudoscalar $\sigma^{\left(1p\right)}$ is in fact proportional to the difference between the squares of the g-tensor principal components.  If the system is magnetically isotropic ($g_{zz}\sim g_{xx} \sim g_{yy}$),  $\sigma^{\left(1p\right)}\sim 0$. If instead it features a large easy-axis magnetic moment $M_0$ ($g_{zz}\sim 2M_0, g_{xx}\sim g_{yy}\sim 0 $), the effect is maximal, with  $\sigma^{\left(1p\right)}\sim \frac{\mu_B^2}{4}\frac{\sin2\chi}{R_I^3}M_0^2 \mu_{\perp}$ , where $\mu_{\perp}$ is the component of $\bm{\bar{\mu}}$ perpendicular to the plane defined by the easy-axis and the probed nucleus.

As discussed in~\cite{Buckingham2014}, in a diamagnetic sample the dominant mechanism responsible for the rotating chiral polarization is related to the temperature-dependent partial orientation of $\bm{\bar{\mu}}$ under the effect of the rotating field $\bm{B}\times\bm{m}^I$. This is driven by the antisymmetric part of the nuclear shielding $\bm{\sigma}$,  which is a generally small second order response property.    While also in paramagnetic molecules partial orientation of $\bm{\bar{\mu}}$ under $\bm{B}\times\bm{m}^I$ is the dominant effect, Eq.~(\ref{sigma1dipolar}) shows that this is now driven by an altogether different mechanism which is itself temperature dependent (hence the overall $\beta^2$ dependence), based on the anisotropy of the first order Zeeman energy of the paramagnet in the strong NMR magnetic field.  In other words, the new mechanism found here is based on the tendency of the molecule to partially align its principal magnetic axis (if any) along the external field.  This is a \emph{first order effect} and thus a much larger orientational driving force than that active in a diamagnetic sample.   

Guided by this insight, in order to estimate $\beta^2\sigma^{\left(1p\right)}$ for promising candidate systems, we considered ten of the Dy$^{3+}$ complexes studied in~\cite{NatComm2013}, which we label here {\bf 1}--{\bf 10} (see Supplementary Material (SM)~\cite{epapsfile} for more information on their chemical structure).
In these systems the ground $^6$H$_{\frac{15}{2}}$ spin-orbit multiplet of Dy$^{3+}$ is split by the low-symmetry crystal field into eight KDs, with a strongly anisotropic ground KD consisting of an almost pure $M_J=\pm15/2$ atomic doublet, thus all excellent candidates to observe the predicted effect. We performed ab initio calculations using \textsc{Molcas} 8.0~\cite{molcas} to determine the g-tensor principal axes, and the electric dipole moment components with respect to such axes, within the well-known CASSCF/RASSI--SO approximation~\cite{Ungur2015}. 
Note that in ionic systems (e.g. complex {\bf 6}) the electric dipole $\bm{\bar{\mu}}$ is origin dependent.  However, since the observable polarization results from the torque applied to the molecule by the rotating field $\bm{B}\times\bm{m}^I$, the detectable $\bm{\bar{\mu}}$ is that computed with respect to the centre of mass.
The results for $\beta^2 \sigma^{\left(1p\right)}$ at 293K for all nuclei of all complexes using Eq.~(\ref{sigma1dipolar}), are reported in Tables XIII--XXII of the SM~\cite{epapsfile}, with details of calculations. 

Our largest estimate for protons at 293K is obtained for complex {\bf 6}~\cite{Chilton2013}, [Dy(paaH*)$_2$(NO$_3$)$_2$(MeOH)]$^+$ (paaH$^*$ = N--(2--pyridyl)acetoacetamide), where for H$_{22}$ we obtain $\beta^2 \sigma^{\left(1p\right)} = 7.33 \cdot 10^6$ ppm au ($1\,\mathrm{ppm}\,\mathrm{au}=10^{-6}\,e\,a_0\,E_h^{-1} \sim 2\cdot10^{-18}\,\mathrm{m}\,\mathrm{V}^{-1}$). In fact {\bf 1}--{\bf 10} all display similar easy-axis anisotropy, but {\bf 6} also features the largest  $\mu_{\perp}\sim20$D.  Note that the largest pseudoscalar obtained in closed-shell molecules for protons is $\sigma^{(1)} = 8.75 \cdot 10^2\,\mathrm{ppm}\,\mathrm{au}$ in (2R)-2-methyloxirane~\cite{Buckingham2015}, which, even assuming a pure chiral liquid, generates an AC-voltage $V$ smaller than the threshold for experimental detection (1$\mu$V), and in a dilute solution (0.1M), it generates $V<1$nV.  Our result is $10^{4}$ times larger, and conservatively assuming a 0.1M solution, and an NMR field of 14.1T, we obtain a chiral polarization $P\sim8.4\,\mathrm{fC}\,\mathrm{m}^{-2}$, and a voltage $V\sim15.2\mu$V, 
well above the 1$\mu$V threshold  (see SM~\cite{epapsfile} for details of calculation of $P$ and $V$).
Higher solubilities, thus higher $P$ and $V$, can be achieved with appropriate choice of ligands, e.g. Na$_9$[Ln(W$_5$O$_{18}$)$_2$] SMMs achieve $>0.1$M in water~\cite{ColetteMichele}, and 
Ln bis-octa(ethyl)tetraazaporphyrin 
are highly soluble in common organic solvents~\cite{Carretta2014}.
 In fact, we find that {\bf 1}--{\bf 10} are all suitable for $^1$H-NMR chiral discrimination, with maximal $P$ ($\,\mathrm{fC}\,\mathrm{m}^{-2}$) and $V$ ($\mu$V), reported as $(P,V)$:  $(1.3, 2.3)$ for H$_{24}$ in {\bf 1}, $(1.9, 3.4)$ for H$_{11}$ in {\bf 2}, $(1.5, 2.7)$ for H$_{15}$ in {\bf 3}, $(1.9, 3.4)$ for H$_1$ in {\bf 4}, 
 $(2.8, 5.0)$ for H$_{8}$ in {\bf 5}, $(1.8, 3.3)$ for H$_{22}$ in {\bf 7}, $(3.4, 6.1)$ for H$_{16}$ in {\bf 8}, $(5.2, 9.3)$ for H$_{20}$ in {\bf 9}, and $(4.2, 7.6)$  for H$_{18}$ in {\bf 10}. Considering more exotic nuclei, the highest value for a diamagnetic system was obtained for $^{125}$Te~\cite{Pelloni2013}, with $\sigma^{(1)} = 8.28 \cdot 10^4\,\mathrm{ppm}\,\mathrm{au}$ (in a 0.1M solution, $V\sim18$ nV), 
while our best result is $\beta^2 \sigma^{\left(1p\right)} = 2.74 \cdot 10^7\,\mathrm{ppm}\,\mathrm{au}$ for $^{17}$O (O$_1$ in complex {\bf 6}).

We conclude by noting that while 3d-complexes are generally less anisotropic than 4f systems, low-coordination environments can trigger unquenched orbital angular momentum, as recently discussed for SMMs based on e.g. Fe$^{+}$~\cite{Long2013iron1}.  Suitable candidates for detection of the chiral polarization could also be e.g. chiral versions of trigonal bis-trispyrazolylborate chelates of Co$^{2+}$ having $g_{\parallel}^2-g_{\perp}^2\sim71$~\cite{Novikov2014,Tierney2012,Jesson1966}, only 5.6 times smaller than in the Dy$^{3+}$ complexes shown here, where $g_{\parallel}^2-g_{\perp}^2\sim400$.
Since our proposal specifically addresses chiral molecules with strong magnetic anisotropy, it could provide an alternative route to {\it direct} chiral discrimination {\it in solution} for these specific systems when other techniques such as circular dichroism are not viable (e.g. if ligands have no chromophores).  Despite the specificity of the proposed approach, we note that synthesis and characterization of enantiomerically pure magnetically anisotropic complexes is strategic in several research areas, among which enantioselective catalysis~\cite{LanthanideReview}, indirect chiral discrimination via NMR using chiral Ln$^{3+}$ shift reagents ~\cite{LanthanideReview}, and stereoselective binding of DNA chains aimed at inhibiting particular biological functions~\cite{LanthanideDNA}.  
We further note that the method could extend the current scope of Ln$^{3+}$ shift reagents, to include {\it direct} chiral discrimination of chiral analytes, via the use of {\it achiral} Ln$^{3+}$ shift reagents. 

We hope these results will encourage experimentalists to gauge our predictions, and design the first experiment to achieve chiral discrimination in pNMR spectroscopy.

We thank Dr C. Boskovic and Dr M. Vonci (University of Melbourne), Prof. K. S. Murray and Prof. G. B. Deacon (Monash University), for useful discussions on the solubility of Ln-SMMs.   A.S. acknowledges support from the Australian Research Council, grant ID: DP150103254


%


\end{document}


\title{Room temperature chiral discrimination in paramagnetic NMR spectroscopy: Supplementary Material}

\author{Alessandro Soncini}
\email{asoncini@unimelb.edu.au}
\author{Simone Calvello}

\affiliation{School of Chemistry, University of Melbourne, VIC3010, Australia}

\maketitle

\subsection{Computational details}
\noindent All calculations were performed with the \textsc{molcas-8.0} program~\cite{molcas}.

\textit{Computational model.} As probe molecules, we chose ten Dy$^{3+}$ complexes with strongly anisotropic easy-axis ground states (see Table~\ref{complex} for names, 
using the experimental geometries from Ref.~\cite{NatComm2013} and reported below in Tables~\ref{geometry1}--\ref{geometry10}. Crystallographic data for these molecules show that they do not possess symmetry (C$_1$ group). 

\begin{center}
\begin{tabular}{cc}
 & Compound \\
\hline
{\bf 1  } & [Dy(acac)$_3$(H$_2$O)$_2$]\\
{\bf 2  } & [Dy(acac)$_3$(dppz)] \\
{\bf 3  } & [Dy(acac)$_3$(dpq)]\\
{\bf 4  } & [Dy(acac)$_3$(phen)]\\
{\bf 5  } & [Dy(hfac)$_3$(dme)]\\
{\bf 6  } & [Dy(paaH*)$_2$(NO$_3$)$_2$(MeOH)]$^+$\\
{\bf 7  } & [Dy(tfpb)$_3$(dppz)]\\
{\bf 8  } & [Dy(tta)$_3$(bipy)]\\
{\bf 9  } & [Dy(tta)$_3$(phen)]\\
{\bf 10 } & [Dy(tta)$_3$(pinene-bipy)]\\
\end{tabular}
\captionof{table}{The ten Dy complexes {\bf 1}--{\bf 10} for which ab initio calculation of the chiral polarization was carried out. Acronyms used above: acac = acetylacetonate, dppz = dipyridophenazine, dpq = dipyridoquinoxaline, phen = 1,10-phenanthroline, hfac = hexafluoroacetylacetone, dme = dimethoxyethane, paaH* = N-(2-Pyridyl)acetoacetamide, tfpb = 4,4,4-trifluoro-1-phenyl-1,3-butandionate, tta = tetradecylthioacetate, bipy = 2,20-bipyridine, pinene-bipy = 4,5-pinene bipyridine.}\label{complex}
\end{center}

\textit{Basis sets.} ANO-rcc basis sets from the \textsc{molcas-8} basis set library are used with the following contractions: VTZP on Dy, VTZ on N and O and VDZ on other nuclei.

\textit{CASSCF.} The active space consists of the seven valence $4f$ orbitals of the Dy$^{3+}$ ion, occupied by 9 electrons. For a given spin, CASSCF calculations are averaged over all the roots that derive from the $4f^n$ electron configuration. For Dy$^{3+}$ we optimised the average energy of the 21 CASSCF roots with $S=5/2$ (from the free ion terms $^6$H, $^6$F, $^6$P), in accordance with previous calculations~\cite{NatComm2013}.

\textit{RASSI.} In this step the spin-orbit coupling Hamiltonian was diagonalized in the basis of the optimized CASSCF wave functions to yield the twofold degenerate ground state. Electric dipole components were also calculated. The module SINGLE\_ANISO~\cite{Ungur2015} has been used to compute the principal magnetic axes and related g-tensor principal components.

All calculations have been performed using the Dysprosium atom as the origin of the coordinate system. Since the rotating electric polarization formula requires all properties to be calculated at the center of mass, the electric dipole moment $\bm{\mu}$ (expressed in a.u.) calculated in the former coordinates has to be translated into the center of mass using the formula:
\begin{equation}
\label{translation}
\bm{\mu'}=\bm{\mu} - Q \mathbf{C_{mass}}
\end{equation}
where Q is the total charge of the molecule and $\mathbf{C_{mass}}$ is the coordinate of the center of mass expressed in atomic units.

\begin{center}
\begin{tabular}{c|ccc|ccc}
Compound & \multicolumn{3}{c|}{CM Coordinates} & \multicolumn{3}{c}{Dipole Moment} \\
 & x & y & z & x & y & z \\
\hline
{\bf 1 } & -0.735 & 0.588 & -0.078 & 1.997 & -0.756 & 0.226 \\
{\bf 2 } & 2.058 & -1.826 & -1.538 & 1.173 & -1.536 & -1.426 \\
{\bf 3 } & 1.425 & 1.790 & 0.547 & 1.726 & 2.182 & 0.433 \\
{\bf 4 } & 0.937 & -0.264 & 0.860 & 2.188 & -0.409 & 2.132 \\
{\bf 5 } & -0.687 & -0.379 & 0.690 & 3.442 & 2.597 & -2.092 \\
{\bf 6 } & 0.611 & -0.442 & 0.147 & 7.511 & -0.739 & -1.755 \\
{\bf 7 } & 0.072 & -0.913 & 0.174 & -1.280 & -2.357 & -1.497 \\
{\bf 8 } & 0.632 & -0.032 & 0.371 & 5.925 & -1.077 & -3.770 \\
{\bf 9 } & -0.662 & -0.239 & 0.219 & -7.391 & 0.034 & -0.735 \\
{\bf 10} & -0.871 & 0.414 & -0.310 & -1.532 & 7.161 & -1.903 \\
\end{tabular}
\captionof{table}{Center of mass (CM) coordinates, in a.u., and dipole moment $\bm{\mu'}$, expressed in a.u., translated (for the ionic molecule {\bf 6}) in the center of mass for the studied compounds.}\label{properties}
\end{center}

\subsection{Pseudoscalar and voltage}
Using Eq. P22, the paramagnetic contribution to the pseudoscalar, $\beta^2 \sigma^{\left(1p\right)}$, calculated at T = 293 K, was computed for all non-lanthanide atoms of the ten probe complexes, and reported below in Tables~\ref{pseudo1}--\ref{pseudo10}. The rotating polarization and the induced voltage were calculated using the following formulae~\cite{BuckinghamFischer2006,Pelloni2013}:
\begin{equation}
\label{volt}
V^I = P^I_y \frac{d}{(\epsilon - 1) \epsilon_0} = \frac{N^I (\beta^2 \sigma^{\left(1p\right)}) (\hbar \gamma_I B_z)^2 I^I(I^I + 1)}{3kT} \frac{d}{(\epsilon - 1) \epsilon_0}
\end{equation}
where $\epsilon$ is the dielectric constant of the medium, $\epsilon_0$ is the permittivity of free space, and $d$ is the separation between the two plates of the capacitor sandwiching the NMR pick-up coil for the chiral polarization detection. For the nucleus I $\gamma_I$ is the gyromagnetic ratio, $I^I$ is the nuclear spin and $N^I = M \cdot NA$, where M is the molarity and NA is Avogadro's number, is the number density, approximable to $N^I \approx 10^{28}$m$^{-3}$ for a pure chiral liquid~\cite{BuckinghamFischer2006}. We assumed a 0.1M solution of our complex in an appropriate solvent, as a conservative estimate of a suitable concentration obtainable in practice (see text for further comments on solubility), yielding a number density of $N^I= 6.022 \cdot 10^{25}$m$^{-3}$. 
In the calculations performed, the ratio $\frac{d}{(\epsilon - 1)}$ was set equal to 0.016 m and $B_z$ was set to 14.1 T, as suggested in previous works~\cite{BuckinghamFischer2006,Pelloni2013}.

The highest value of $\beta^2 \sigma^{\left(1p\right)}$ obtained for the hydrogen atoms, which constitute the easiest and the most frequently measured NMR nucleus, consists of $7.33 \cdot 10^{6}$ ppm au for H$_{22}$ in complex {\bf 6}, $10^{4}$ times larger than the largest value obtained for diamagnetic systems, in (2R)-2-dimethyloxirane~\cite{Buckingham2015}. Including all the nuclei, the highest value increases to $2.74 \cdot 10^7$ ppm au for an oxygen atom (O$_1$) in complex {\bf 6}, to be compared with the highest previous estimate of $8.28 \cdot 10^4$ ppm au for a tellurium nucleus in 1,2-M-ditellurin~\cite{Buckingham2015}.

The maximum voltage we obtained consists of 15.2 $\mu$V, induced by the resonant precession of H$_{22}$ in complex {\bf 6} (see table~\ref{pseudo6}), well above the experimental threshold of 1$\mu$V.

\section{Geometries}
\begin{small}
\begin{center}
\begin{tabular}{c|ccc|c|ccc|c|ccc}
Atom & x & y & z & Atom & x & y & z & Atom & x & y & z \\
\hline
Dy  &  0.00000 &  0.00000 &  0.00000 & C1  & -1.46448 & -3.94277 &  2.05137 & H1  & -0.68899 & -4.00038 &  2.61444 \\
O1  & -0.66534 & -1.90736 &  1.18518 & C2  & -1.17322 & -3.02569 &  0.88540 & H2  & -1.68698 & -4.81693 &  1.72281 \\
O2  & -1.06398 & -1.42002 & -1.51292 & C3  & -1.50151 & -3.43635 & -0.39120 & H3  & -2.20169 & -3.59310 &  2.55691 \\
O3  & -0.76256 &  0.82638 &  2.09716 & C4  & -1.44490 & -2.62318 & -1.51845 & H4  & -1.78026 & -4.31627 & -0.50514 \\
O4  & -2.02054 &  1.11207 & -0.34927 & C5  & -1.87094 & -3.19417 & -2.85478 & H5  & -2.74477 & -2.86794 & -3.07935 \\
O5  &  1.08251 &  2.00888 &  0.37000 & C6  & -1.15415 &  2.47835 &  3.73505 & H6  & -1.89110 & -4.15336 & -2.80168 \\
O6  &  0.21421 &  1.09558 & -2.10487 & C7  & -1.37365 &  1.89584 &  2.37107 & H7  & -1.24523 & -2.92556 & -3.53180 \\
\end{tabular}
\captionof{table}{Experimental geometry, used in calculations, for complex {\bf 1}. All coordinates are expressed in \AA.}\label{geometry1}
\end{center}
\end{small}
\begin{small}
\begin{center}
\begin{tabular}{c|ccc|c|ccc|c|ccc}
Atom & x & y & z & Atom & x & y & z & Atom & x & y & z \\
\hline    
Dy  &  0.00000 &  0.00000 &  0.00000 & C1  &  1.93717 & -2.65581 & -1.17994 & H1  &  5.88651 & -7.12522 & -2.60195 \\
O1  & -1.32985 &  1.41244 &  1.21864 & C2  &  5.96298 & -6.50784 & -3.29283 & H2  &  7.49440 & -5.97818 & -6.08490 \\
O2  &  0.17925 & -0.63848 &  2.21859 & C3  &  5.19834 & -5.35245 & -3.27205 & H3  & -2.80444 &  2.10705 & -2.99414 \\
O3  &  2.19456 &  0.39502 &  0.61607 & C4  &  2.60130 & -3.78727 & -1.21760 & H4  & -4.14306 &  1.29725 & -2.90324 \\
O4  & -1.36991 &  0.44785 & -1.80848 & C5  &  6.93650 & -5.80512 & -5.36026 & H5  & -3.03104 &  0.86254 & -3.91877 \\
O5  &  0.51316 &  2.17971 & -0.71113 & C6  & -3.20044 &  1.23820 & -3.04998 & H6  &  6.37014 & -4.09271 & -6.00698 \\
O6  & -1.83849 & -1.41503 &  0.16804 & C7  &  2.83080 &  0.39631 &  1.68226 & H7  &  3.66050 & -0.97368 & -4.94210 \\
\end{tabular}
\captionof{table}{Experimental geometry, used in calculations, for complex {\bf 2}. All coordinates are expressed in \AA.}\label{geometry2}
\end{center}
\end{small}
\begin{small}
\begin{center}
\begin{tabular}{c|ccc|c|ccc|c|ccc}
Atom & x & y & z & Atom & x & y & z & Atom & x & y & z \\
\hline         
Dy  &  0.00000 &  0.00000 &  0.00000 & C1  &  3.40693 &  0.06819 &  0.68587 & H1  &  3.26932 & -0.83997 &  0.53907 \\
O1  &  0.77990 &  1.00205 & -1.93607 & C2  &  4.68842 &  0.50894 &  0.95107 & H2  &  5.39802 & -0.09191 &  0.94475 \\
O2  &  1.36124 & -1.55616 & -1.03631 & C3  &  4.90786 &  1.80228 &  1.21468 & H3  &  5.77029 &  2.10488 &  1.38990 \\
O3  & -1.38343 & -1.79498 & -0.39226 & C4  &  3.82396 &  2.70016 &  1.22731 & H4  &  6.08778 &  6.16473 &  2.54697 \\
O4  & -1.91156 &  0.85823 & -1.01421 & C5  &  3.97367 &  4.09205 &  1.57143 & H5  &  4.28069 &  7.53053 &  2.46331 \\
O5  & -1.40924 &  0.43114 &  1.85715 & C6  &  5.27268 &  5.80920 &  2.27388 & H6  &  0.52778 &  6.18288 &  1.14207 \\
O6  &  0.82386 & -1.21070 &  1.78927 & C7  &  4.16817 &  6.64152 &  2.21548 & H7  & -1.49287 &  5.26496 &  0.46645 \\
\end{tabular}
\captionof{table}{Experimental geometry, used in calculations, for complex {\bf 3}. All coordinates are expressed in \AA.}\label{geometry3}
\end{center}
\end{small}
\begin{small}
\begin{center}
\begin{tabular}{c|ccc|c|ccc|c|ccc}
Atom & x & y & z & Atom & x & y & z & Atom & x & y & z \\
\hline   
Dy  &  0.00000 &  0.00000 &  0.00000 & C1  &  3.36302 &  0.71987 & -0.00923 & H1  &  3.07425 &  1.24945 & -0.71571 \\
O1  &  0.56613 &  2.24850 &  0.10270 & C2  &  4.73756 &  0.63161 &  0.23499 & H2  &  5.34397 &  1.08764 & -0.30432 \\
O2  & -1.28528 &  0.93065 &  1.67658 & C3  &  5.16680 & -0.12283 &  1.25400 & H3  &  6.07936 & -0.20269 &  1.41681 \\
O3  &  0.98498 &  0.39393 & -2.06179 & C4  &  4.25485 & -0.78270 &  2.07241 & H4  &  5.52754 & -1.67163 &  3.41559 \\
O4  & -1.66828 &  0.87854 & -1.37712 & C5  &  4.62529 & -1.56446 &  3.21498 & H5  &  3.99020 & -2.62992 &  4.74278 \\
O5  &  0.99363 & -2.02406 & -0.59215 & C6  &  3.71364 & -2.14447 &  3.99850 & H6  &  1.53730 & -3.05652 &  5.27918 \\
O6  & -1.65046 & -1.63234 &  0.10270 & C7  &  2.31028 & -2.03309 &  3.70923 & H7  & -0.65229 & -2.72028 &  4.77766 \\
\end{tabular}
\captionof{table}{Experimental geometry, used in calculations, for complex {\bf 4}. All coordinates are expressed in \AA.}\label{geometry4}
\end{center}
\end{small}
\begin{small}
\begin{center}
\begin{tabular}{c|ccc|c|ccc|c|ccc}
Atom & x & y & z & Atom & x & y & z & Atom & x & y & z \\
\hline 
Dy  &  0.00000 &  0.00000 &  0.00000 & C1  &  3.34275 & -0.50471 &  0.57232 & H1  &  3.46536 & -1.34257 &  0.08135 \\
O1  &  2.39513 &  0.32214 & -0.12253 & C2  &  2.88909 &  0.78989 & -1.38533 & H2  &  3.00644 & -0.70000 &  1.47106 \\
O2  &  0.71885 &  1.75691 & -1.55593 & C3  &  2.14816 &  1.99472 & -1.72028 & H3  &  4.19928 & -0.03537 &  0.63890 \\
O3  & -1.02399 &  1.94275 &  0.75540 & C4  & -0.02383 &  2.92236 & -1.87215 & H4  &  2.75421 &  0.10007 & -2.08227 \\
O4  &  0.73812 &  0.39143 &  2.18463 & C5  & -1.20616 &  2.46721 &  1.87465 & H5  &  3.85772 &  0.98834 & -1.32084 \\
O5  & -1.72638 & -0.85907 &  1.31503 & C6  & -0.64389 &  2.10182 &  3.08337 & H6  &  2.33908 &  2.25459 & -2.65646 \\
O6  &  0.69783 & -2.08280 &  0.76788 & C7  &  0.30197 &  1.07811 &  3.13954 & H7  &  2.43367 &  2.73652 & -1.12944 \\
\end{tabular}
\captionof{table}{Experimental geometry, used in calculations, for complex {\bf 5}. All coordinates are expressed in \AA.}\label{geometry5}
\end{center}
\end{small}
\begin{small}
\begin{center}
\begin{tabular}{c|ccc|c|ccc|c|ccc}
Atom & x & y & z & Atom & x & y & z & Atom & x & y & z \\
\hline
Dy  &  0.00000 &  0.00000 &  0.00000 & C1  & -0.92578 &  3.18610 & -0.43022 & H1  &  0.88620 & -5.13818 & -0.48570 \\
O1  &  0.22937 & -2.16529 & -0.76376 & C2  & -2.05447 &  4.07625 & -0.91653 & H2  & -1.84580 &  4.40676 & -1.81520 \\
O2  &  1.30069 &  1.78059 &  0.79342 & C3  &  1.22593 &  3.03949 &  0.72825 & H3  & -2.15844 &  4.83521 & -0.30533 \\
O3  & -1.06221 &  1.94542 & -0.60636 & C4  &  0.89333 & -3.18263 & -0.38689 & H4  & -2.88806 &  3.56165 & -0.94311 \\
O4  &  1.42870 &  0.40897 & -1.84273 & C5  &  4.25215 &  4.16407 &  2.58473 & H5  &  4.14321 &  5.10500 &  2.51207 \\
O5  & -2.44002 & -0.47744 & -0.15612 & C6  &  3.90801 & -2.03931 &  1.62108 & H6  &  4.54196 & -1.42120 &  1.20104 \\
O6  &  2.04553 & -0.80958 &  0.83064 & C7  &  2.59018 & -1.96520 &  0.90797 & H7  &  4.25794 & -2.95309 &  1.56669 \\
\end{tabular}
\captionof{table}{Experimental geometry, used in calculations, for complex {\bf 6}. All coordinates are expressed in \AA.}\label{geometry6}
\end{center}
\end{small}
\begin{scriptsize}
\begin{center}
\begin{tabular}{c|ccc|c|ccc|c|ccc}
Atom & x & y & z & Atom & x & y & z & Atom & x & y & z \\
\hline               
Dy  &  0.00000 &  0.00000 &  0.00000 & C1  &  0.95173 & -3.14104 & -0.96504 & H1  & -2.06903 &  3.76571 &  0.88426 \\
O1  & -0.24787 &  2.22212 &  0.52584 & C2  &  1.72933 & -4.21655 & -1.40402 & H2  & -2.65413 & -5.63267 &  0.07636 \\
O2  & -1.49778 &  0.08861 &  1.78206 & C3  & -0.39485 & -3.36208 & -0.46536 & H3  & -1.93618 &  2.25066 & -3.58489 \\
O3  &  0.42373 &  0.71555 & -2.17923 & C4  & -0.91946 & -4.65184 & -0.38713 & H4  & -3.86108 & -3.76995 &  0.61691 \\
O4  &  2.12505 &  0.91313 &  0.14174 & C5  & -0.19218 &  1.37075 & -3.05602 & H5  &  3.18558 & -8.80409 & -2.17806 \\
O5  & -2.05235 &  0.43826 & -1.00590 & C6  &  2.55243 &  1.96518 &  0.64843 & H6  & -1.17301 & -9.52769 & -0.55409 \\
O6  &  0.98339 & -1.04767 &  1.82643 & C7  & -0.08195 & -5.79293 & -0.76657 & H7  & -3.29294 &  5.47297 &  1.81475 \\
\end{tabular}
\captionof{table}{Experimental geometry, used in calculations, for complex {\bf 7}. All coordinates are expressed in \AA.}\label{geometry7}
\end{center}
\end{scriptsize}
\begin{footnotesize}
\begin{center}
\begin{tabular}{c|ccc|c|ccc|c|ccc}
Atom & x & y & z & Atom & x & y & z & Atom & x & y & z \\
\hline     
Dy  &  0.00000 &  0.00000 &  0.00000 & C1  &  6.17715 &  1.24526 &  2.25054 & H1  &  7.07424 &  1.33300 &  1.94809 \\
O1  &  1.98459 &  0.62446 &  1.06400 & C2  &  5.80861 &  1.40253 &  3.54282 & H2  &  6.43980 &  1.61773 &  4.21963 \\
O2  & -0.56902 &  0.51189 &  2.16594 & C3  &  4.40941 &  1.22539 &  3.84104 & H3  &  3.98265 &  1.30320 &  4.68494 \\
O3  &  0.80980 & -1.92326 &  0.99632 & C4  &  3.78096 &  0.89430 &  2.54030 & H4  &  1.85630 &  0.43574 &  4.23232 \\
O4  & -1.84916 & -1.39186 &  0.38466 & C5  &  2.37049 &  0.66420 &  2.26534 & H5  &  4.23608 & -4.35512 &  4.08215 \\
O5  &  0.89905 &  1.64422 & -1.36785 & C6  &  1.48694 &  0.49865 &  3.35881 & H6  &  2.27653 & -5.14643 &  5.21370 \\
O6  & -1.64743 &  1.62601 & -0.35306 & C7  &  0.12810 &  0.42084 &  3.22345 & H7  &  0.17874 & -4.33360 &  4.20694 \\
\end{tabular}
\captionof{table}{Experimental geometry, used in calculations, for complex {\bf 8}. All coordinates are expressed in \AA.}\label{geometry8}
\end{center}
\end{footnotesize}
\begin{footnotesize}
\begin{center}
\begin{tabular}{c|ccc|c|ccc|c|ccc}
Atom & x & y & z & Atom & x & y & z & Atom & x & y & z \\
\hline            
Dy  &  0.00000 &  0.00000 &  0.00000 & C1  & -3.42833 & -2.57014 & -5.02550 & H1  & -4.32653 & -2.83832 & -5.18116 \\
O1  & -0.79991 & -0.88383 & -2.00300 & C2  & -2.44798 & -2.64218 & -5.96550 & H2  & -2.61998 & -2.98610 & -6.83447 \\
O2  &  1.73433 &  0.06792 & -1.53754 & C3  & -1.12534 & -2.17027 & -5.58618 & H3  & -0.33897 & -2.12107 & -6.11662 \\
O3  & -0.90087 & -2.07916 &  0.47392 & C4  & -1.31864 & -1.76578 & -4.09457 & H4  &  1.26068 & -1.24447 & -4.42251 \\
O4  &  1.62800 & -1.19372 &  1.19328 & C5  & -0.37032 & -1.18783 & -3.13644 & H5  & -4.22264 & -6.06726 &  0.35151 \\
O5  & -0.81348 &  0.29703 &  2.14083 & C6  &  0.98132 & -0.95786 & -3.55959 & H6  & -2.83365 & -6.95021 &  2.13479 \\
O6  &  1.19085 &  1.74699 &  0.95752 & C7  &  1.86538 & -0.33419 & -2.73747 & H7  & -0.97407 & -5.46102 &  2.71510 \\
\end{tabular}
\captionof{table}{Experimental geometry, used in calculations, for complex {\bf 9}. All coordinates are expressed in \AA.}\label{geometry9}
\end{center}
\end{footnotesize}
\begin{footnotesize}
\begin{center}
\begin{tabular}{c|ccc|c|ccc|c|ccc}
Atom & x & y & z & Atom & x & y & z & Atom & x & y & z \\
\hline 
Dy  &  0.00000 &  0.00000 &  0.00000 & C1  &  3.06851 &  0.79179 &  1.03378 & H1  &  5.75045 &  0.94581 &  2.08534 \\
O1  &  2.10599 &  0.98194 &  0.21859 & C2  &  4.12948 &  1.81099 &  1.12325 & H2  &  6.87864 &  3.15155 &  1.91965 \\
O2  &  1.38353 & -1.61931 &  0.98186 & C3  &  5.39117 &  1.70831 &  1.68880 & H3  &  5.59188 &  4.76592 &  0.61735 \\
O3  &  0.54858 &  0.83172 & -2.12865 & C4  &  6.04103 &  2.96901 &  1.55845 & H4  &  3.79612 & -0.40045 &  2.57909 \\
O4  & -0.12979 & -1.80756 & -1.54764 & C5  &  5.28637 &  3.91215 &  0.82612 & H5  & -0.88624 &  1.01426 & -6.34374 \\
O5  &  0.10345 &  1.04089 &  2.12952 & C6  &  3.16867 & -0.38524 &  1.89314 & H6  & -0.52409 &  3.39684 & -6.85737 \\
O6  & -1.51369 & -1.12873 &  1.43474 & C7  &  2.38627 & -1.41775 &  1.71089 & H7  &  0.65440 &  4.80205 & -5.28776  \\
\end{tabular}
\captionof{table}{Experimental geometry, used in calculations, for complex {\bf 10}. All coordinates are expressed in \AA.}\label{geometry10}
\end{center}
\end{footnotesize}

\section{Computational Results}
\begin{small} 
\begin{center}
\begin{tabular}{c|ccc|c|ccc|c|ccc}
Atom & $\beta^2 \sigma^{\left(1p\right)}$ & $|$P$|$ & $|$V$|$ & Atom & $\beta^2 \sigma^{\left(1p\right)}$ & $|$P$|$ & $|$V$|$ & Atom & $\beta^2 \sigma^{\left(1p\right)}$ & $|$P$|$ & $|$V$|$ \\
\hline
O1  & -5.24$\cdot 10^6$ & 1.29$\cdot 10^{-15}$ &  2.325 & C1 & -5.13$\cdot 10^5$ & 3.72$\cdot 10^{-17}$ &  0.067 & H1 & -6.26$\cdot 10^5$ & 7.16$\cdot 10^{-16}$ &  1.295 \\ 
O2 &  8.52$\cdot 10^6$ & 2.09$\cdot 10^{-15}$ &  3.782 & C2 & -2.94$\cdot 10^5$ & 2.13$\cdot 10^{-17}$ &  0.039 & H2 & -1.54$\cdot 10^5$ & 1.76$\cdot 10^{-16}$ &  0.318 \\ 
O3 & -2.63$\cdot 10^5$ & 6.46$\cdot 10^{-17}$ &  0.117 & C3 &  1.30$\cdot 10^6$ & 9.42$\cdot 10^{-17}$ &  0.170 & H3 & -6.27$\cdot 10^5$ & 7.19$\cdot 10^{-16}$ &  1.298 \\ 
O4 & -2.11$\cdot 10^4$ & 5.17$\cdot 10^{-18}$ &  0.009 & C4 &  3.00$\cdot 10^6$ & 2.17$\cdot 10^{-16}$ &  0.392 & H4 &  6.84$\cdot 10^5$ & 7.84$\cdot 10^{-16}$ &  1.416 \\ 
O5 &  6.30$\cdot 10^6$ & 1.55$\cdot 10^{-15}$ &  2.795 & C5 &  1.12$\cdot 10^6$ & 8.09$\cdot 10^{-17}$ &  0.146 & H5 &  8.19$\cdot 10^5$ & 9.38$\cdot 10^{-16}$ &  1.694 \\ 
O6 & -8.19$\cdot 10^6$ & 2.01$\cdot 10^{-15}$ &  3.634 & C6 &  4.26$\cdot 10^5$ & 3.09$\cdot 10^{-17}$ &  0.056 & H6 &  7.65$\cdot 10^5$ & 8.76$\cdot 10^{-16}$ &  1.583 \\ 
O7 &  1.11$\cdot 10^5$ & 2.72$\cdot 10^{-17}$ &  0.049 & C7 &  9.29$\cdot 10^5$ & 6.72$\cdot 10^{-17}$ &  0.122 & H7 &  9.49$\cdot 10^5$ & 1.09$\cdot 10^{-15}$ &  1.965 \\
\end{tabular}
\captionof{table}{Values of $\beta^2 \sigma^{\left(1p\right)}$, expressed in ppm au, of the rotating electric polarization, expressed in $\frac{C}{m^2}$, and of the induced voltage, expressed in $\mu$V, for non-lanthanide atoms of complex {\bf 1}. 1 ppm au of $\beta^2 \sigma^{\left(1p\right)}$ = 1.94469097 $\cdot 10^{-18} \frac{m}{V}$.}\label{pseudo1}
\end{center}
\end{small}
\begin{small}
\begin{center}
\begin{tabular}{c|ccc|c|ccc|c|ccc}
Atom & $\beta^2 \sigma^{\left(1p\right)}$ & $|$P$|$ & $|$V$|$ & Atom & $\beta^2 \sigma^{\left(1p\right)}$ & $|$P$|$ & $|$V$|$ & Atom & $\beta^2 \sigma^{\left(1p\right)}$ & $|$P$|$ & $|$V$|$ \\
\hline    
O1 & -2.11$\cdot 10^4$ & 5.18$\cdot 10^{-18}$ &  0.009 & C1 &  1.92$\cdot 10^5$ & 1.39$\cdot 10^{-17}$ &  0.025 & H1 &  1.49$\cdot 10^4$ & 1.71$\cdot 10^{-17}$ &  0.031 \\
O2 &  9.43$\cdot 10^6$ & 2.32$\cdot 10^{-15}$ &  4.183 & C2 &  8.06$\cdot 10^3$ & 5.83$\cdot 10^{-19}$ &  0.001 & H2 & -2.65$\cdot 10^3$ & 3.04$\cdot 10^{-18}$ &  0.005 \\
O3 & -6.06$\cdot 10^6$ & 1.49$\cdot 10^{-15}$ &  2.687 & C3 &  5.95$\cdot 10^3$ & 4.31$\cdot 10^{-19}$ &  0.001 & H3 &  1.11$\cdot 10^6$ & 1.27$\cdot 10^{-15}$ &  2.292 \\
O4 &  7.34$\cdot 10^6$ & 1.80$\cdot 10^{-15}$ &  3.258 & C4 &  1.32$\cdot 10^5$ & 9.58$\cdot 10^{-18}$ &  0.017 & H4 &  3.85$\cdot 10^5$ & 4.41$\cdot 10^{-16}$ &  0.796 \\
O5 & -1.35$\cdot 10^6$ & 3.31$\cdot 10^{-16}$ &  0.599 & C5 & -2.59$\cdot 10^3$ & 1.88$\cdot 10^{-19}$ &  \textless 0.001 & H5 &  6.78$\cdot 10^5$ & 7.77$\cdot 10^{-16}$ &  1.403 \\
O6 & -1.02$\cdot 10^7$ & 2.50$\cdot 10^{-15}$ &  4.513 & C6 &  8.34$\cdot 10^5$ & 6.04$\cdot 10^{-17}$ &  0.109 & H6 & -5.96$\cdot 10^3$ & 6.82$\cdot 10^{-18}$ &  0.012 \\
N1 &  1.48$\cdot 10^6$ & 2.36$\cdot 10^{-17}$ &  0.043 & C7 & -7.64$\cdot 10^5$ & 5.53$\cdot 10^{-17}$ &  0.100 & H7 &  2.88$\cdot 10^4$ & 3.30$\cdot 10^{-17}$ &  0.060 \\
\end{tabular}
\captionof{table}{Values of $\beta^2 \sigma^{\left(1p\right)}$, expressed in ppm au, of the rotating electric polarization, expressed in $\frac{C}{m^2}$, and of the induced voltage, expressed in $\mu$V, for non-lanthanide atoms of complex {\bf 2}. 1 ppm au of $\beta^2 \sigma^{\left(1p\right)}$ = 1.94469097 $\cdot 10^{-18} \frac{m}{V}$.}\label{pseudo2}
\end{center}
\end{small}
\begin{small}
\begin{center}
\begin{tabular}{c|ccc|c|ccc|c|ccc}
Atom & $\beta^2 \sigma^{\left(1p\right)}$ & $|$P$|$ & $|$V$|$ & Atom & $\beta^2 \sigma^{\left(1p\right)}$ & $|$P$|$ & $|$V$|$ & Atom & $\beta^2 \sigma^{\left(1p\right)}$ & $|$P$|$ & $|$V$|$ \\
\hline         
O1  &  2.16$\cdot 10^6$ & 5.31$\cdot 10^{-16}$ &  0.959 & C1 & -4.55$\cdot 10^5$ & 3.29$\cdot 10^{-17}$ &  0.059 & H1 & -4.39$\cdot 10^5$ & 5.03$\cdot 10^{-16}$ &  0.909 \\
O2 & -9.22$\cdot 10^6$ & 2.26$\cdot 10^{-15}$ &  4.091 & C2 & -1.86$\cdot 10^5$ & 1.34$\cdot 10^{-17}$ &  0.024 & H2 & -1.45$\cdot 10^5$ & 1.67$\cdot 10^{-16}$ &  0.301 \\
O3 &  3.98$\cdot 10^4$ & 9.76$\cdot 10^{-18}$ &  0.018 & C3 & -9.78$\cdot 10^4$ & 7.08$\cdot 10^{-18}$ &  0.013 & H3 & -6.42$\cdot 10^4$ & 7.36$\cdot 10^{-17}$ &  0.133 \\
O4 &  5.56$\cdot 10^6$ & 1.36$\cdot 10^{-15}$ &  2.466 & C4 & -6.00$\cdot 10^4$ & 4.34$\cdot 10^{-18}$ &  0.008 & H4 & -2.17$\cdot 10^3$ & 2.49$\cdot 10^{-18}$ &  0.004 \\
O5 & -9.70$\cdot 10^6$ & 2.38$\cdot 10^{-15}$ &  4.303 & C5 & -9.71$\cdot 10^3$ & 7.03$\cdot 10^{-19}$ &  0.001 & H5 &  1.55$\cdot 10^3$ & 1.78$\cdot 10^{-18}$ &  0.003 \\
O6 &  1.21$\cdot 10^7$ & 2.96$\cdot 10^{-15}$ &  5.350 & C6 & -1.97$\cdot 10^3$ & 1.43$\cdot 10^{-19}$ &  \textless 0.001 & H6 &  2.33$\cdot 10^4$ & 2.67$\cdot 10^{-17}$ &  0.048 \\
N1 & -7.10$\cdot 10^5$ & 1.13$\cdot 10^{-17}$ &  0.020 & C7 &  1.83$\cdot 10^3$ & 1.32$\cdot 10^{-19}$ &  \textless 0.001 & H7 &  3.88$\cdot 10^4$ & 4.45$\cdot 10^{-17}$ &  0.080 \\
\end{tabular}
\captionof{table}{Values of $\beta^2 \sigma^{\left(1p\right)}$, expressed in ppm au, of the rotating electric polarization, expressed in $\frac{C}{m^2}$, and of the induced voltage, expressed in $\mu$V, for non-lanthanide atoms of complex {\bf 3}. 1 ppm au of $\beta^2 \sigma^{\left(1p\right)}$ = 1.94469097 $\cdot 10^{-18} \frac{m}{V}$.}\label{pseudo3}
\end{center}
\end{small}
\begin{small}
\begin{center}
\begin{tabular}{c|ccc|c|ccc|c|ccc}
Atom & $\beta^2 \sigma^{\left(1p\right)}$ & $|$P$|$ & $|$V$|$ & Atom & $\beta^2 \sigma^{\left(1p\right)}$ & $|$P$|$ & $|$V$|$ & Atom & $\beta^2 \sigma^{\left(1p\right)}$ & $|$P$|$ & $|$V$|$ \\
\hline   
O1  & -1.03$\cdot 10^7$ & 2.53$\cdot 10^{-15}$ &  4.565 & C1 & -8.61$\cdot 10^5$ & 6.23$\cdot 10^{-17}$ &  0.113 & H1 & -1.65$\cdot 10^6$ & 1.89$\cdot 10^{-15}$ &  3.413 \\
O2 &  1.36$\cdot 10^7$ & 3.34$\cdot 10^{-15}$ &  6.030 & C2 & -1.82$\cdot 10^5$ & 1.32$\cdot 10^{-17}$ &  0.024 & H2 & -1.79$\cdot 10^5$ & 2.05$\cdot 10^{-16}$ &  0.371 \\
O3 &  4.48$\cdot 10^6$ & 1.10$\cdot 10^{-15}$ &  1.988 & C3 & -8.56$\cdot 10^3$ & 6.20$\cdot 10^{-19}$ &  0.001 & H3 &  3.26$\cdot 10^3$ & 3.74$\cdot 10^{-18}$ &  0.007 \\
O4 & -1.46$\cdot 10^3$ & 3.58$\cdot 10^{-19}$ &  0.001 & C4 &  4.31$\cdot 10^4$ & 3.12$\cdot 10^{-18}$ &  0.006 & H4 &  1.83$\cdot 10^4$ & 2.10$\cdot 10^{-17}$ &  0.038 \\
O5 &  5.20$\cdot 10^6$ & 1.28$\cdot 10^{-15}$ &  2.309 & C5 &  1.87$\cdot 10^4$ & 1.35$\cdot 10^{-18}$ &  0.002 & H5 & -2.72$\cdot 10^4$ & 3.11$\cdot 10^{-17}$ &  0.056 \\
O6 & -1.17$\cdot 10^7$ & 2.88$\cdot 10^{-15}$ &  5.200 & C6 & -2.61$\cdot 10^4$ & 1.89$\cdot 10^{-18}$ &  0.003 & H6 & -1.21$\cdot 10^5$ & 1.39$\cdot 10^{-16}$ &  0.251 \\
N1 & -7.73$\cdot 10^5$ & 1.23$\cdot 10^{-17}$ &  0.022 & C7 & -1.33$\cdot 10^5$ & 9.61$\cdot 10^{-18}$ &  0.017 & H7 & -2.33$\cdot 10^5$ & 2.67$\cdot 10^{-16}$ &  0.482 \\
\end{tabular}
\captionof{table}{Values of $\beta^2 \sigma^{\left(1p\right)}$, expressed in ppm au, of the rotating electric polarization, expressed in $\frac{C}{m^2}$, and of the induced voltage, expressed in $\mu$V, for non-lanthanide atoms of complex {\bf 4}. 1 ppm au of $\beta^2 \sigma^{\left(1p\right)}$ = 1.94469097 $\cdot 10^{-18} \frac{m}{V}$.}\label{pseudo4}
\end{center}
\end{small}
\begin{small}
\begin{center}
\begin{tabular}{c|ccc|c|ccc|c|ccc}
Atom & $\beta^2 \sigma^{\left(1p\right)}$ & $|$P$|$ & $|$V$|$ & Atom & $\beta^2 \sigma^{\left(1p\right)}$ & $|$P$|$ & $|$V$|$ & Atom & $\beta^2 \sigma^{\left(1p\right)}$ & $|$P$|$ & $|$V$|$ \\
\hline 
O1  & -6.45$\cdot 10^5$ & 1.58$\cdot 10^{-16}$ &  0.286 & C1 &  5.41$\cdot 10^4$ & 3.92$\cdot 10^{-18}$ &  0.007 & H1 &  2.27$\cdot 10^5$ & 2.60$\cdot 10^{-16}$ &  0.469 \\
O2 &  1.46$\cdot 10^6$ & 3.59$\cdot 10^{-16}$ &  0.648 & C2 & -1.30$\cdot 10^6$ & 9.44$\cdot 10^{-17}$ &  0.171 & H2 &  3.07$\cdot 10^4$ & 3.52$\cdot 10^{-17}$ &  0.064 \\
O3 &  5.70$\cdot 10^6$ & 1.40$\cdot 10^{-15}$ &  2.527 & C3 &  1.34$\cdot 10^4$ & 9.67$\cdot 10^{-19}$ &  0.002 & H3 & -1.81$\cdot 10^4$ & 2.07$\cdot 10^{-17}$ &  0.037 \\
O4 & -8.47$\cdot 10^6$ & 2.08$\cdot 10^{-15}$ &  3.758 & C4 &  1.97$\cdot 10^6$ & 1.42$\cdot 10^{-16}$ &  0.257 & H4 & -2.29$\cdot 10^6$ & 2.63$\cdot 10^{-15}$ &  4.748 \\
O5 & -4.20$\cdot 10^6$ & 1.03$\cdot 10^{-15}$ &  1.865 & C5 &  2.41$\cdot 10^5$ & 1.74$\cdot 10^{-17}$ &  0.031 & H5 & -4.24$\cdot 10^5$ & 4.86$\cdot 10^{-16}$ &  0.878 \\
O6 &  9.79$\cdot 10^6$ & 2.40$\cdot 10^{-15}$ &  4.342 & C6 & -1.37$\cdot 10^6$ & 9.92$\cdot 10^{-17}$ &  0.179 & H6 & -4.14$\cdot 10^5$ & 4.74$\cdot 10^{-16}$ &  0.857 \\
O7 & -2.32$\cdot 10^6$ & 5.71$\cdot 10^{-16}$ &  1.031 & C7 & -3.18$\cdot 10^6$ & 2.30$\cdot 10^{-16}$ &  0.416 & H7 &  1.06$\cdot 10^6$ & 1.21$\cdot 10^{-15}$ &  2.191 \\
\end{tabular}
\captionof{table}{Values of $\beta^2 \sigma^{\left(1p\right)}$, expressed in ppm au, of the rotating electric polarization, expressed in $\frac{C}{m^2}$, and of the induced voltage, expressed in $\mu$V, for non-lanthanide atoms of complex {\bf 5}. 1 ppm au of $\beta^2 \sigma^{\left(1p\right)}$ = 1.94469097 $\cdot 10^{-18} \frac{m}{V}$.}\label{pseudo5}
\end{center}
\end{small}
\begin{small}
\begin{center}
\begin{tabular}{c|ccc|c|ccc|c|ccc}
Atom & $\beta^2 \sigma^{\left(1p\right)}$ & $|$P$|$ & $|$V$|$ & Atom & $\beta^2 \sigma^{\left(1p\right)}$ & $|$P$|$ & $|$V$|$ & Atom & $\beta^2 \sigma^{\left(1p\right)}$ & $|$P$|$ & $|$V$|$ \\
\hline
O1  &  2.74$\cdot 10^7$ & 6.72$\cdot 10^{-15}$ &  12.15 & C1 &  6.04$\cdot 10^4$ & 4.37$\cdot 10^{-18}$ &  0.008 & H1 &  1.32$\cdot 10^6$ & 1.51$\cdot 10^{-15}$ &  2.734 \\
O2 &  1.92$\cdot 10^7$ & 4.71$\cdot 10^{-15}$ &  8.504 & C2 & -9.18$\cdot 10^5$ & 6.65$\cdot 10^{-17}$ &  0.120 & H2 & -1.46$\cdot 10^6$ & 1.67$\cdot 10^{-15}$ &  3.018 \\
O3 & -1.23$\cdot 10^7$ & 3.03$\cdot 10^{-15}$ &  5.470 & C3 &  7.23$\cdot 10^6$ & 5.23$\cdot 10^{-16}$ &  0.946 & H3 &  1.63$\cdot 10^5$ & 1.87$\cdot 10^{-16}$ &  0.338 \\
O4 & -2.46$\cdot 10^6$ & 6.05$\cdot 10^{-16}$ &  1.093 & C4 &  5.03$\cdot 10^6$ & 3.64$\cdot 10^{-16}$ &  0.658 & H4 & -1.36$\cdot 10^6$ & 1.56$\cdot 10^{-15}$ &  2.821 \\
O5 & -2.21$\cdot 10^6$ & 5.42$\cdot 10^{-16}$ &  0.980 & C5 &  6.96$\cdot 10^5$ & 5.04$\cdot 10^{-17}$ &  0.091 & H5 &  6.64$\cdot 10^5$ & 7.61$\cdot 10^{-16}$ &  1.374 \\
O6 & -1.86$\cdot 10^7$ & 4.58$\cdot 10^{-15}$ &  8.273 & C6 & -2.42$\cdot 10^6$ & 1.75$\cdot 10^{-16}$ &  0.316 & H6 & -1.58$\cdot 10^6$ & 1.81$\cdot 10^{-15}$ &  3.275 \\
O7 & -4.83$\cdot 10^6$ & 1.19$\cdot 10^{-15}$ &  2.143 & C7 & -5.51$\cdot 10^6$ & 3.99$\cdot 10^{-16}$ &  0.721 & H7 & -1.42$\cdot 10^6$ & 1.63$\cdot 10^{-15}$ &  2.939 \\
\end{tabular}
\captionof{table}{Values of $\beta^2 \sigma^{\left(1p\right)}$, expressed in ppm au, of the rotating electric polarization, expressed in $\frac{C}{m^2}$, and of the induced voltage, expressed in $\mu$V, for non-lanthanide atoms of complex {\bf 6}. 1 ppm au of $\beta^2 \sigma^{\left(1p\right)}$ = 1.94469097 $\cdot 10^{-18} \frac{m}{V}$.}\label{pseudo6}
\end{center}
\end{small}
\begin{scriptsize}
\begin{center}
\begin{tabular}{c|ccc|c|ccc|c|ccc}
Atom & $\beta^2 \sigma^{\left(1p\right)}$ & $|$P$|$ & $|$V$|$ & Atom & $\beta^2 \sigma^{\left(1p\right)}$ & $|$P$|$ & $|$V$|$ & Atom & $\beta^2 \sigma^{\left(1p\right)}$ & $|$P$|$ & $|$V$|$ \\
\hline            
O1  & -1.22$\cdot 10^6$ & 3.00$\cdot 10^{-16}$ &  0.542 & C1 & -2.60$\cdot 10^5$ & 1.89$\cdot 10^{-17}$ &  0.034 & H1 & -3.91$\cdot 10^5$ & 4.47$\cdot 10^{-16}$ &  0.808 \\
O2 &  1.24$\cdot 10^7$ & 3.04$\cdot 10^{-15}$ &  5.500 & C2 & -8.26$\cdot 10^4$ & 5.98$\cdot 10^{-18}$ &  0.011 & H2 &  2.45$\cdot 10^4$ & 2.81$\cdot 10^{-17}$ &  0.051 \\
O3 &  6.69$\cdot 10^6$ & 1.64$\cdot 10^{-15}$ &  2.968 & C3 & -3.42$\cdot 10^5$ & 2.47$\cdot 10^{-17}$ &  0.045 & H3 & -1.08$\cdot 10^6$ & 1.23$\cdot 10^{-15}$ &  2.230 \\
O4 & -9.17$\cdot 10^5$ & 2.25$\cdot 10^{-16}$ &  0.407 & C4 & -1.07$\cdot 10^5$ & 7.78$\cdot 10^{-18}$ &  0.014 & H4 &  2.00$\cdot 10^5$ & 2.30$\cdot 10^{-16}$ &  0.415 \\
O5 & -1.12$\cdot 10^7$ & 2.76$\cdot 10^{-15}$ &  4.987 & C5 &  1.67$\cdot 10^5$ & 1.21$\cdot 10^{-17}$ &  0.022 & H5 & -2.65$\cdot 10^4$ & 3.04$\cdot 10^{-17}$ &  0.055 \\
O6 & -8.51$\cdot 10^6$ & 2.09$\cdot 10^{-15}$ &  3.775 & C6 & -4.11$\cdot 10^5$ & 2.97$\cdot 10^{-17}$ &  0.054 & H6 & -2.07$\cdot 10^4$ & 2.37$\cdot 10^{-17}$ &  0.043 \\
N1 &  2.66$\cdot 10^5$ & 4.24$\cdot 10^{-18}$ &  0.008 & C7 & -1.04$\cdot 10^5$ & 7.51$\cdot 10^{-18}$ &  0.014 & H7 & -5.08$\cdot 10^4$ & 5.81$\cdot 10^{-17}$ &  0.105 \\
\end{tabular}
\captionof{table}{Values of $\beta^2 \sigma^{\left(1p\right)}$, expressed in ppm au, of the rotating electric polarization, expressed in $\frac{C}{m^2}$, and of the induced voltage, expressed in $\mu$V, for non-lanthanide atoms of complex {\bf 7}. 1 ppm au of $\beta^2 \sigma^{\left(1p\right)}$ = 1.94469097 $\cdot 10^{-18} \frac{m}{V}$.}\label{pseudo7}
\end{center}
\end{scriptsize}
\begin{footnotesize}
\begin{center}
\begin{tabular}{c|ccc|c|ccc|c|ccc}
Atom & $\beta^2 \sigma^{\left(1p\right)}$ & $|$P$|$ & $|$V$|$ & Atom & $\beta^2 \sigma^{\left(1p\right)}$ & $|$P$|$ & $|$V$|$ & Atom & $\beta^2 \sigma^{\left(1p\right)}$ & $|$P$|$ & $|$V$|$ \\
\hline     
O1  &  1.94$\cdot 10^6$ & 4.77$\cdot 10^{-16}$ &  0.862 & C1 &  2.13$\cdot 10^4$ & 1.54$\cdot 10^{-18}$ &  0.003 & H1 &  5.10$\cdot 10^4$ & 5.84$\cdot 10^{-17}$ &  0.105 \\
O2 & -8.17$\cdot 10^6$ & 2.00$\cdot 10^{-15}$ &  3.622 & C2 & -1.02$\cdot 10^5$ & 7.41$\cdot 10^{-18}$ &  0.013 & H2 & -8.25$\cdot 10^4$ & 9.45$\cdot 10^{-17}$ &  0.171 \\
O3 & -1.30$\cdot 10^7$ & 3.19$\cdot 10^{-15}$ &  5.773 & C3 & -3.52$\cdot 10^5$ & 2.55$\cdot 10^{-17}$ &  0.046 & H3 & -4.32$\cdot 10^5$ & 4.95$\cdot 10^{-16}$ &  0.894 \\
O4 &  1.89$\cdot 10^7$ & 4.65$\cdot 10^{-15}$ &  8.399 & C4 & -5.22$\cdot 10^5$ & 3.78$\cdot 10^{-17}$ &  0.068 & H4 & -2.28$\cdot 10^6$ & 2.61$\cdot 10^{-15}$ &  4.720 \\
O5 &  1.29$\cdot 10^6$ & 3.16$\cdot 10^{-16}$ &  0.571 & C5 & -2.43$\cdot 10^6$ & 1.76$\cdot 10^{-16}$ &  0.318 & H5 & -8.28$\cdot 10^5$ & 9.48$\cdot 10^{-16}$ &  1.713 \\
O6 & -1.08$\cdot 10^7$ & 2.64$\cdot 10^{-15}$ &  4.774 & C6 & -3.95$\cdot 10^6$ & 2.86$\cdot 10^{-16}$ &  0.517 & H6 & -6.83$\cdot 10^5$ & 7.82$\cdot 10^{-16}$ &  1.413 \\
N1 & -7.13$\cdot 10^6$ & 1.14$\cdot 10^{-16}$ &  0.205 & C7 & -5.64$\cdot 10^6$ & 4.09$\cdot 10^{-16}$ &  0.738 & H7 & -9.57$\cdot 10^5$ & 1.10$\cdot 10^{-15}$ &  1.980 \\
\end{tabular}
\captionof{table}{Values of $\beta^2 \sigma^{\left(1p\right)}$, expressed in ppm au, of the rotating electric polarization, expressed in $\frac{C}{m^2}$, and of the induced voltage, expressed in $\mu$V, for non-lanthanide atoms of complex {\bf 8}. 1 ppm au of $\beta^2 \sigma^{\left(1p\right)}$ = 1.94469097 $\cdot 10^{-18} \frac{m}{V}$.}\label{pseudo8}
\end{center}
\end{footnotesize}
\begin{footnotesize}
\begin{center}
\begin{tabular}{c|ccc|c|ccc|c|ccc}
Atom & $\beta^2 \sigma^{\left(1p\right)}$ & $|$P$|$ & $|$V$|$ & Atom & $\beta^2 \sigma^{\left(1p\right)}$ & $|$P$|$ & $|$V$|$ & Atom & $\beta^2 \sigma^{\left(1p\right)}$ & $|$P$|$ & $|$V$|$ \\
\hline        
O1  &  3.52$\cdot 10^6$ & 8.64$\cdot 10^{-16}$ &  1.561 & C1 &  2.55$\cdot 10^5$ & 1.85$\cdot 10^{-17}$ &  0.033 & H1 &  2.06$\cdot 10^5$ & 2.36$\cdot 10^{-16}$ &  0.427 \\ 
O2 & -1.30$\cdot 10^7$ & 3.20$\cdot 10^{-15}$ &  5.780 & C2 &  1.34$\cdot 10^5$ & 9.73$\cdot 10^{-18}$ &  0.018 & H2 &  8.52$\cdot 10^4$ & 9.75$\cdot 10^{-17}$ &  0.176 \\ 
O3 &  6.14$\cdot 10^6$ & 1.51$\cdot 10^{-15}$ &  2.723 & C3 &  4.00$\cdot 10^4$ & 2.90$\cdot 10^{-18}$ &  0.005 & H3 & -9.11$\cdot 10^4$ & 1.04$\cdot 10^{-16}$ &  0.189 \\ 
O4 & -9.23$\cdot 10^6$ & 2.27$\cdot 10^{-15}$ &  4.094 & C4 &  3.84$\cdot 10^5$ & 2.78$\cdot 10^{-17}$ &  0.050 & H4 & -7.40$\cdot 10^5$ & 8.47$\cdot 10^{-16}$ &  1.530 \\ 
O5 & -1.42$\cdot 10^7$ & 3.48$\cdot 10^{-15}$ &  6.284 & C5 &  5.01$\cdot 10^2$ & 3.63$\cdot 10^{-20}$ &  \textless 0.001 & H5 &  3.30$\cdot 10^5$ & 3.78$\cdot 10^{-16}$ &  0.683 \\ 
O6 &  2.31$\cdot 10^7$ & 5.67$\cdot 10^{-15}$ &  10.25 & C6 & -1.60$\cdot 10^6$ & 1.16$\cdot 10^{-16}$ &  0.209 & H6 &  5.79$\cdot 10^4$ & 6.63$\cdot 10^{-17}$ &  0.120 \\ 
N1 & -6.12$\cdot 10^6$ & 9.76$\cdot 10^{-17}$ &  0.176 & C7 & -4.22$\cdot 10^6$ & 3.06$\cdot 10^{-16}$ &  0.552 & H7 & -3.58$\cdot 10^5$ & 4.10$\cdot 10^{-16}$ &  0.740 \\
\end{tabular}
\captionof{table}{Values of $\beta^2 \sigma^{\left(1p\right)}$, expressed in ppm au, of the rotating electric polarization, expressed in $\frac{C}{m^2}$, and of the induced voltage, expressed in $\mu$V, for non-lanthanide atoms of complex {\bf 9}. 1 ppm au of $\beta^2 \sigma^{\left(1p\right)}$ = 1.94469097 $\cdot 10^{-18} \frac{m}{V}$.}\label{pseudo9}
\end{center}
\end{footnotesize}
\begin{footnotesize}
\begin{center}
\begin{tabular}{c|ccc|c|ccc|c|ccc}
Atom & $\beta^2 \sigma^{\left(1p\right)}$ & $|$P$|$ & $|$V$|$ & Atom & $\beta^2 \sigma^{\left(1p\right)}$ & $|$P$|$ & $|$V$|$ & Atom & $\beta^2 \sigma^{\left(1p\right)}$ & $|$P$|$ & $|$V$|$ \\
\hline 
O1  & -4.71$\cdot 10^6$ & 1.16$\cdot 10^{-15}$ &  2.087 & C1 &  2.62$\cdot 10^6$ & 1.90$\cdot 10^{-16}$ &  0.343 & H1 &  4.62$\cdot 10^5$ & 5.29$\cdot 10^{-16}$ &  0.956 \\
O2 &  6.28$\cdot 10^6$ & 1.54$\cdot 10^{-15}$ &  2.787 & C2 &  5.93$\cdot 10^5$ & 4.29$\cdot 10^{-17}$ &  0.078 & H2 &  1.37$\cdot 10^5$ & 1.57$\cdot 10^{-16}$ &  0.284 \\
O3 & -7.83$\cdot 10^6$ & 1.92$\cdot 10^{-15}$ &  3.475 & C3 &  4.05$\cdot 10^5$ & 2.93$\cdot 10^{-17}$ &  0.053 & H3 & -5.27$\cdot 10^4$ & 6.03$\cdot 10^{-17}$ &  0.109 \\
O4 &  1.60$\cdot 10^7$ & 3.92$\cdot 10^{-15}$ &  7.088 & C4 &  1.58$\cdot 10^5$ & 1.14$\cdot 10^{-17}$ &  0.021 & H4 &  2.61$\cdot 10^6$ & 2.99$\cdot 10^{-15}$ &  5.402 \\
O5 &  1.87$\cdot 10^7$ & 4.59$\cdot 10^{-15}$ &  8.300 & C5 & -6.45$\cdot 10^3$ & 4.67$\cdot 10^{-19}$ &  0.001 & H5 &  7.83$\cdot 10^5$ & 8.97$\cdot 10^{-16}$ &  1.621 \\
O6 & -2.63$\cdot 10^7$ & 6.46$\cdot 10^{-15}$ &  11.68 & C6 &  4.25$\cdot 10^6$ & 3.08$\cdot 10^{-16}$ &  0.556 & H6 &  1.91$\cdot 10^5$ & 2.19$\cdot 10^{-16}$ &  0.396 \\
N1 & -3.66$\cdot 10^5$ & 5.84$\cdot 10^{-18}$ &  0.011 & C7 &  5.03$\cdot 10^6$ & 3.65$\cdot 10^{-16}$ &  0.659 & H7 & -1.73$\cdot 10^5$ & 1.98$\cdot 10^{-16}$ &  0.358 \\
\end{tabular}
\captionof{table}{Values of $\beta^2 \sigma^{\left(1p\right)}$, expressed in ppm au, of the rotating electric polarization, expressed in $\frac{C}{m^2}$, and of the induced voltage, expressed in $\mu$V, for non-lanthanide atoms of complex {\bf 10}. 1 ppm au of $\beta^2 \sigma^{\left(1p\right)}$ = 1.94469097 $\cdot 10^{-18} \frac{m}{V}$.}\label{pseudo10}
\end{center}
\end{footnotesize}


%